\def\selectedoptions{final}

\documentclass[\selectedoptions]{aipproc}

\layoutstyle{8x11double}
\usepackage{fix2col}
\begin{document}

\title 
      [THE FREE-FREE OPACITY IN WARM, DENSE, AND WEAKLY IONIZED HELIUM]
      {THE FREE-FREE OPACITY IN WARM, DENSE, AND WEAKLY IONIZED HELIUM}

\keywords{Dense Matter, Helium, Pressure Ionization, Quantum: Molecuar Dynamics}
\classification{51.70.+f, 64.30.+t, 71.15.-m, 72.20.-i, 72.80.-r, 78.40.Dw}

\author{Piotr M. Kowalski}{
  address={Vanderbilt University, Nashville, TN 37235},
  altaddress={X-7, MS-F699, LANL, Los Alamos, NM 87545},
  email={kowalski@lanl.gov},
}

\iftrue
\author{Stephane Mazevet}{
  address={T-4, MS-B283, LANL, Los Alamos, NM 87545},
  email={smazevet@lanl.gov},
}
\author{Didier Saumon}{
  address={X-7, MS-F699, LANL, Los Alamos, NM 87545},
  email={dsaumon@lanl.gov},
}
\fi

\copyrightyear  {2005}

\begin{abstract}
We investigate the ionization and the opacity of warm, dense helium under conditions found in the atmospheres of cool white dwarf stars. 
Our particular interest is in densities up to $\rm 3 \ g/cm^{3}$ and temperatures from 1000K to 10000K. For these physical conditions various approaches
for modeling the ionization equilibrium predict ionization fractions that differ by orders of magnitudes. Furthermore, estimates of the density 
at which helium pressure-ionizes vary from $\rm 0.3$ to $\rm 14 \ g/cm^{3}$. In this context, the value of the electron-atom inverse bremsstrahlung 
absorption is highly uncertain. We present new results obtained from a non-ideal chemical model for
the ionization equilibrium, from Quantum Molecular Dynamics (QMD) simulations, and from the analysis of experimental data to better understand 
the ionization fraction in fluid helium in the weak ionization limit.\\
\end{abstract}

\date{\today}

\maketitle

\section{INTRODUCTION}

We are interested in the opacity of $\rm He$ and $\rm He/H$ mixtures for densities up to $\rm 3 \ g/cm^{3}$ and temperatures of $T\rm=1000K$ to $\rm 10000K$ 
to model the atmospheres of very cool white dwarf stars. Our work is motivated by the fact that there exists a wide range 
of predictions for the density 
\vskip 3pt
\begin{table}[h]
\begin{tabular}{ccc}
\hline
\tablehead{1}{c}{b}{Refference} & \tablehead{1}{c}{b}{
$\rho_{PI}$ ($g/cm^{3}$)} & \tablehead{1}{c}{b}{Source} 
\\
\hline
\cite{Ebeling} & 2 & chem \\
\cite{Fontaine} & 0.3 & chem \\
\cite{Redmer} &  4 & chem \\
\cite{Winisdoerffer} &  >6.5 & chem \\
\cite{Young} & 7.5-14 & LMTO \\
\cite{Fortov} & $\sim 1.5-2$ & chem \& exp \\
\hline
\end{tabular}
\caption{
Density $\rho_{PI}$ for the pressure ionization of helium from experiments (exp), chemical models (chem) and a $T\rm=0 \ K$ quantum mechanical 
calculation (LMTO)}
\end{table}
at which helium pressure-ionizes (Table 1).
This translates into large uncertainties in the number 
of free electrons and the value of the free-free opacity in the regime of our interest. We approach this problem by constructing more reliable models for ionization equilibrium of $\rm He$ 
on the basis of a chemical model, QMD simulations, and available experimental data \cite{Fortov}.

\section{The current state of modeling}

There is a large span in predictions for the density at which the pressure ionization phenomenon occurs in helium (Table 1.). 

For the chemical models \cite{Fortov}-\cite{Winisdoerffer}, this arises from oversimplified treatments of the physics and the presence of free parameters. 
The interactions between
atoms and ions are usually described by a polarization potential with a hard sphere cut off at short distance (e.g. \cite{Winisdoerffer}), 
which varies between models. Often they are simply neglected (see \cite{Redmer}). 
In particular, It is not correct to treat the $\rm e-He$ interaction with a polarization potential \cite{Geltman}, as we discuss below.

The more sophisticated quantum calculations of the $T\rm=0 \ K$ solid predict a high density for pressure ionization \cite{Young}, which we confirm 
with QMD simulations. On the other hand the only published experimental data on helium for the regime of pressure ionization \cite{Fortov} 
suggest that this phenomenon occurs at density $\rm \sim1.5-2.0 \ g/cm^{3}$.

\section{The electron-helium interaction}

The energy of a free electron in cryogenic helium has been shown experimentally to be as large as $V_{e}\rm \sim +1.5 \ eV$ (positive) at densities 
of $\rm \sim 0.2 \ g/cm^{3}$ \cite{Broomall}. This is in the quantitative agreement 
with the prediction of the Lenz potential energy \cite{Tankersley}
\begin{equation} V_{e}=2\pi\hbar^{2}n(\rm He \it )a/m_{e},\end{equation}
where $\rm a \sim 1.3 \ a.u.$ is the scattering length \cite{Tankersley}, and $n(\rm He \it )$ is the number density of helium.
This is because the $\rm e-He$ interaction is strongly repulsive for $\rm r>1 \ a.u.$ \cite{Kestner}, and significantly different from the polarization 
interaction \cite{Geltman} (Fig. 1a, solid \& dotted curves). The quantum mechanical calculations for the energy of an electron in dense helium, using the potential 
of \cite{Kestner}, are in good agreement with experimental data \cite{Boltjes}. This shows that free electrons interact strongly with the atoms in dense $\rm He$, but not 
through an attractive polarization potential. This is an important element in chemical models.

To introduce the interaction of quantum mechanical electrons with the atoms in a chemical model of the thermodynamics, we performed Density Functional Theory (DFT)
calculation of the free energy of free electron in dense $\rm He$. The results are well approximated by the Lenz formula (Fig. 1b), which we adopt in the chemical model. 
 
\begin{figure}
\includegraphics[width=14.8cm]{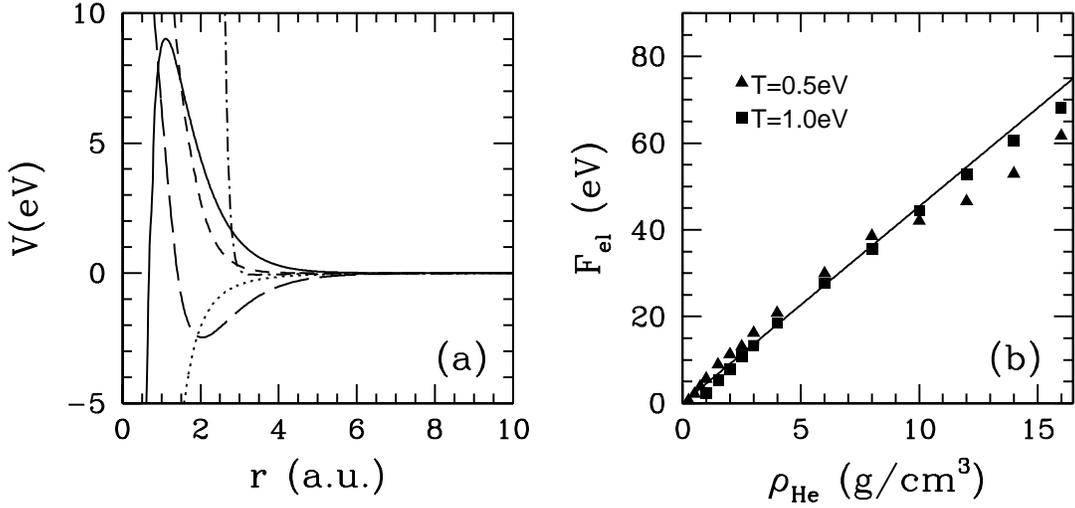}
\caption{(a) The interaction potentials for the chemical model. The lines represent the following potential curves: $\rm e-He$ of \cite{Kestner} (solid) and \cite{Geltman} (dotted),
$\rm He-He$ of \cite{Ross} (dashed), $\rm He-He^{+}$ of \cite{Cencek} (long dashed), and $\rm He-He_{2}^{+}$ of \cite{Scifoni} (dash-dotted). (b) The DFT free energy of a free electron in helium. 
The line represents the Lenz potential energy (Eq. 1).}
\label{fig:OldPhase} 
\end{figure}
 
\section{The chemical model for the ionization equilibrium}

We constructed a chemical model considering the following species: $\rm He,He^+,He_{2}^{+},e^{-}$.
The interactions between atoms and ions are described by the potentials shown in Fig. 1a. 
In the chemical picture, the non-ideal effects can be treated as a shift in 
the ionization/dissociation energies \cite{Fontaine}. Considering the reactions $\rm He \leftrightarrow He^{+}+e^{-}$ and $\rm He_{2}^{+}\leftrightarrow He+He^{+}$ the shifts are $\Delta I_{1}=\mu^{nid}(\rm He \it)-\mu^{nid}(\rm He^{+}\it)-\mu^{nid}(e)$ and 
$\Delta I_{2}=\mu^{nid}(\rm He_{2}^{+} \it)-\mu^{nid}(\rm He \it)-\mu^{nid}(\rm He^{+}\it)$. The non-ideal contributions to the chemical potentials for the $\rm He$ atom 
and the ions where obtained through the numerical solution of the Ornstein-Zernike equation in the Percus-Yevick approximation \cite{Martynow}. This approach is valid in the low ionization limit. 
The results are shown in Fig. 2a, 
where they are compared with the shift in the ionization energy extracted from conductivity data \cite{Fortov} through the relation $\sigma_{exp}/\sigma_{id}=e^{\Delta I_{1}/k_{B}T}$,
where $\sigma_{id}$ is the ideal gas conductivity. 
The agreement with the experiment is good. In this model $\rm He$ pressure-ionizes at $\rm \sim 2 \ g/cm^{3}$ because of strong attractive $\rm He-He^{+}$ interaction (Fig. 1a), 
which favors ionization.
This model is questionable, however, because this $\rm He-He^{+}$ potential inevitably leads to a bound state ($\rm He^{+}_{2}$), which would greatly reduce the interaction with
the other neighboring $\rm He$ atoms.
\begin{figure}
\includegraphics[width=14.8cm]{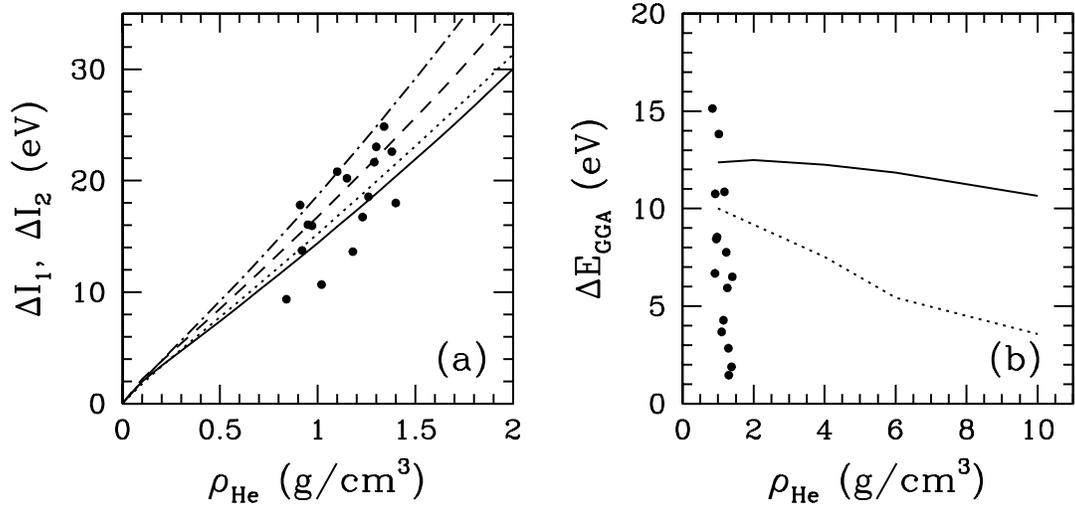}
\caption{(a) The change in the ionization/dissociation potential predicted by the chemical model. The lines represent the results for 
$\Delta I_{1}$ (solid, dotted) and $\Delta I_{2}$ (dashed, dash-dotted) for $T\rm=1.0 \ eV$ and $T\rm=2.0 \ eV$ respectively. The filled circles represent the value of $\Delta I_{1}^{exp}$ 
extracted from the experimental data of \cite{Fortov}. (b) The QMD band gap for $T\rm=0.5 \ eV$ (solid line) and $T\rm=1.5 \ eV$ (dotted line). The filled circles represent the effective value of 
ionization energy $\rm 24.6 \ eV -$$\Delta I_{1}^{exp}$.}
\label{fig:OldPhase} 
\end{figure}

\section{Quantum Molecular Dynamics results}

We also conducted QMD-DFT calculations of dense helium, using the Viena Ab-initio simulation package. The results in terms of band gap (ionization energy) is presented on Figure 2b. 
We find that in this model, helium pressure-ionizes above a density of $\rm 10 \ g/cm^{3}$. 
The resulting conductivities at the experimental conditions are 2 orders of magnitude smaller and show a very strong temperature dependence rather 
than the strong density dependence reported in \cite{Fortov}. 
It is well known that band gaps are underestimated using GGA functional. More accurate functionals would only increase the gap and worsen the disagreement
with the experimental data.

\begin{figure}
\includegraphics[width=7.0cm]{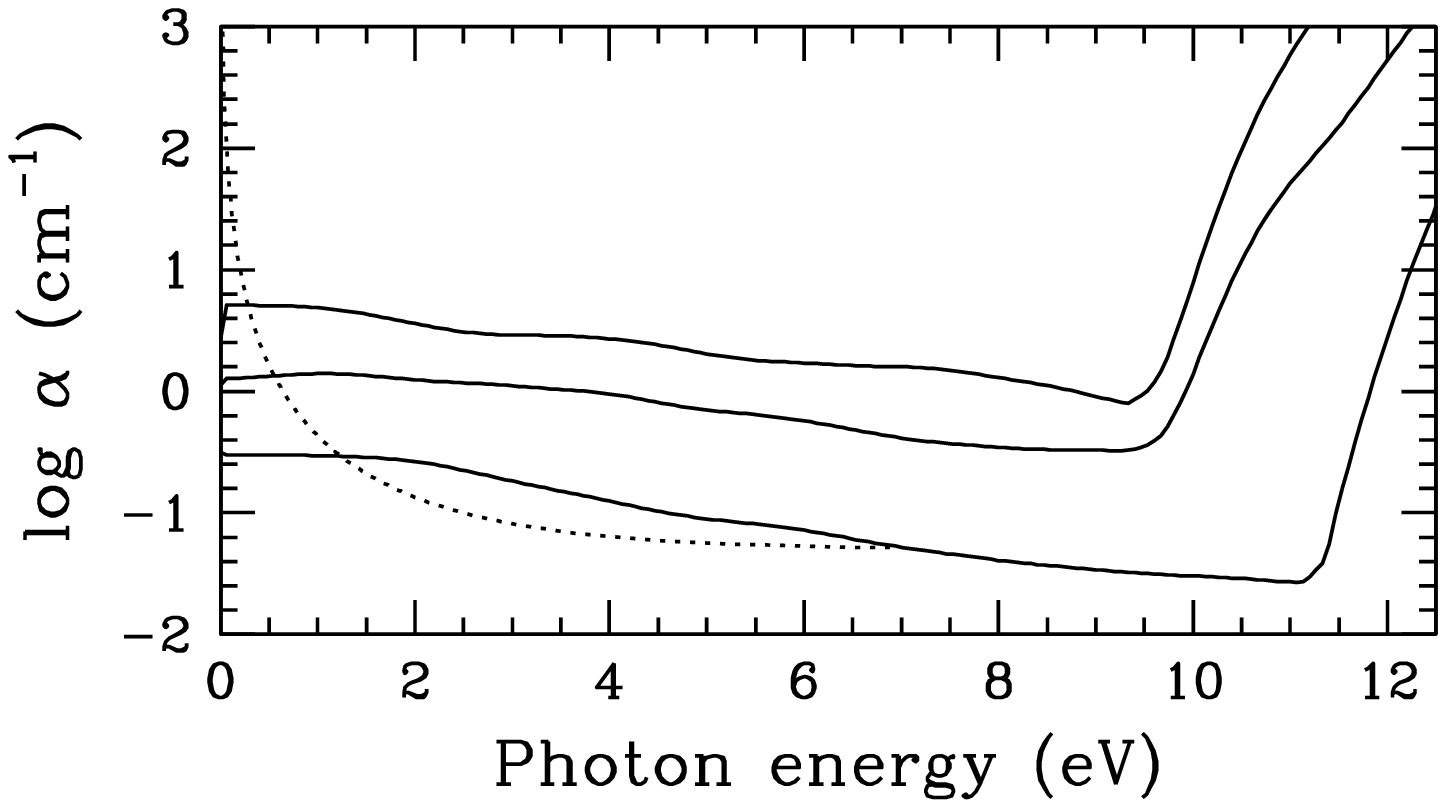}
\caption{The absorption coefficient of He obtained from QMD simulations (solid lines) for temperature $\rm T=0.5 \ eV$ and densities  
$\rho\rm =0.5,2.0$ and $\rm 4.0 \ g/cm^{3}$ (from the bottom to top). The dotted line represents the standard free-free absorption coefficient for $\rho \rm =0.5 \ g/cm^{3}$ 
\cite{Iglesias}, adopting the density of free electrons of the QMD simulation.}
\label{fig:OldPhase} 
\end{figure}

\section{The free-free opacity from helium}

The free-free (inverse bremsstrahlung) opacity of dense helium is determined by the ionization fraction and $\rm e-He$ collisions.
In a dense medium, the collisions can be described with the classical Drude model \cite{Jackson}. 
On Figure 3 we compare the QMD absorption coefficient \cite{Mazevet} with the standard free-free absorption coefficient corrected for the correlations in dense fluid but not 
the $\rm e-He$ collision frequency \cite{Iglesias}. For the astrophysical applications, we are interested in the spectral region with photon energies from zero to $\rm 4 \ eV$. 
The standard free-free frequency behaviour is erroreous, because the absorption processes for photons with small energies are driven by the frequent $\rm e-He$ collisions
rather than the slowly varying electric field of radiation.  

\section{Conclusions}

Our goal is to calculate the free-free absorption from dense, non-ideal, weakly ionized helium. This source of opacity depends on the ionization 
fraction and the frequency of $\rm e-He$ collisions. We use two different approaches to solve for the number density of free electrons in dense helium: 
a chemical model and QMD-DFT simulations. In the first method we have introduced a new description of the $\rm e-He$ interaction and use 
ab initio potentials for the He-ions interactions. This gives us a good agreement with the experiment, but the result is driven by a questionable $\rm He-He^{+}$ potential. 
On the other hand, the more sophisticated QMD simulations suggest that pressure ionization of $\rm He$ does not occur below $\rm 10 \ g/cm^{3}$, which is inconsistent 
with the conductivity data. This large discrepancy between models and the conductivity measurements is astrophysically significant. New experiments that probe the pressure
ionization of dense $\rm He$ are essential in finding a resolution of this problem.  

\begin{theacknowledgments}
We are grateful to Vladimir E. Fortov for providing conductivity data
in dense helium. This work was supported by the U.S. Department of Energy under contract W-7405-ENG-36.
\end{theacknowledgments}

\end{document}